\documentstyle[epsf]{mn}

\def\plotone#1{\centering \leavevmode
\epsfxsize=9.0truecm \epsfbox{#1}}
\def\plottwo#1#2{{\centering \leavevmode
\epsfxsize=8.5truecm \epsfbox{#1}} \centering \leavevmode
\epsfxsize=8.5truecm \epsfbox{#2}}

\begin{document}

\title[The X--ray Spectrum of SS Cyg]
{The X--ray Spectrum of the Dwarf Nova SS Cyg in quiescence and outburst}
\author[C. Done and J.P. Osborne]
{C. Done$^{1}$ and J. P. Osborne$^2$ \\
$^1$Department of Physics, University of Durham, South Road, Durham, DH1 3LE\\ 
$^2$Department of Physics \& Astronomy, University of Leicester, University Road, Leicester. LE1 7RH.}

\maketitle

\begin{abstract}

We reanalyse archival GINGA and ASCA X--ray data from SS Cyg in both
quiescence and outburst, using multi--temperature plasma models for
the continuum and line emission, together with their reflection from
the X--ray illuminated white dwarf and accretion disk.  Reflection is
clearly detected in all the spectra, but its contribution is larger in
the softer X--ray spectra seen in outburst than in quiescence. This
supports models in which the quiescent inner disk is not present or
not optically thick, so that the only reflector is the white dwarf
surface rather than the white dwarf plus the disk. The amount of
reflection in outburst is also more consistent with the hard X--rays
forming a corona over the white dwarf surface rather than just an
equatorial band.  We detect partially ionized absorption in the ASCA
outburst spectrum, which is probably the X--ray signature of the
outflowing wind. The ASCA data also allow a detailed measure of the
elemental abundances. We find that all detectable lines from the hot
plasma (except perhaps Si and S) are a factor $\sim 2.5\times$ weaker
than expected from solar abundances.  We examine possible deviations
from coronal equilibrium, but conclude that the heavy elements are
truly underabundant in SS Cyg.

In quiescence the underlying intrinsic spectrum is consistent with a
single temperature plasma. This conflicts with the expected cooling of
the material, and requires that either the plasma is reheated or that
the cooler emission is masked by absorption in an optically thin inner
disk which affects only that part of the boundary layer emission below
the disk. Such absorption is also required if the data are to match
the theoretical models (Narayan \& Popham 1993) for the quiescent hard
X--ray emission. By contrast, the outburst spectrum is much softer and
is dominated by the cooling components, so that it cannot be fit by a
single temperature model. There are no readily available (or
believable) theoretical models for the hard X--ray emission in
outburst, but the emission observed can be described by material with
a continuous temperature distribution. Although the outburst spectrum
is much softer than the quiescent spectrum, the total luminosity 
of the X--ray component in
both cases is similar, showing that the optically thin plasma emission
is important even in outburst. 

\end{abstract}
 
 
\section{INTRODUCTION}

The non--magnetic Cataclysmic Variable stars are systems where a white
dwarf accretes via a disk from a Roche lobe filling, low mass,
companion star. Dwarf novae (DN) are a subclass of these systems which
show optical/UV outbursts of 2--6 magnitudes which are thought to be
triggered by a disk instability giving a transient increase
in the accretion rate (see e.g. the reviews by Cordova 1995, Warner
1995 and Osaki 1996). As the accreting material falls inwards, the
transition from the the disk velocity field to that of the white dwarf
will lead to up to half of the accretion energy being dissipated
(Pringle 1977, Kluzniak 1987).

Detailed theoretical models of the boundary layer emission are not
well developed due to the inherent complexity of modelling the
strongly shearing, probably turbulent transition region between the
disk and white dwarf surface.  However, general characteristics of the
emission can be derived. Material leaves the inner edge of the disk
with a temperature $\sim 10^5$ K, and then is heated (presumably by shocks)
as it joins the boundary layer. If the boundary layer is
optically thick then the gas cools efficiently and so collapses quickly
onto the white dwarf surface, forming a geometrically thin equatorial
belt around the white dwarf.  The luminosity emerges at the local
blackbody temperature, which is of order $20-50$ eV for the case where
all the remaining accretion luminosity is radiated (Pringle 1977; see
also Frank King and Raine 1992).  Conversely, if the gas is optically
thin then radiative cooling is much less efficient, the gas can be
heated up to much higher temperatures before reaching equilibrium with
radiative cooling and/or adiabatic expansion, giving
temperatures of order $\sim 10$ keV. Since
the optical depth of the accreting gas is strongly linked to the mass
accretion rate this predicts that mainly EUVE/soft X--rays are
expected from DN in outburst, while hard X--rays can be produced by DN
in quiescence (Pringle and Savonije 1979; Tylenda 1981; King and
Shaviv 1985, Narayan and Popham 1993; Popham and Narayan 1995). 
A logical extension of this
model is that the vertical density gradients in the disk could ensure
that some fraction of the accreting material remains optically thin
even in outburst, producing some hard X--ray emission, as observed
(Patterson and Raymond 1985a; 1985b).

The resulting hard X--ray spectrum in both quiescence and outburst
should certainly consist of multi--temperature components as the
accreting gas must cool in order to settle onto the white dwarf
photosphere.  Thus the continuum should be characterised by continuous
temperature distribution, with the contribution from each
temperature component being weighted by its emissivity. 
However, in addition to
this intrinsic hard X--ray continuum, there should also be a component
formed from reflection of this spectrum from the illuminated white
dwarf surface and/or accretion disk.  The reflection albedo is energy
dependent, as at low energies the scattering probability is decreased
by photo--electric absorption, whereas at higher energies Compton
downscattering and the reduction of the scattering cross--section
deplete the number of photons reflected. This results in a broad
spectral ``bump'', centered around 20 keV, while the dependence on
photo--electric absorption at low energies gives a pronounced feature
at the iron K edge energy in the reflected spectrum and an associated
Fe $K\alpha$ fluorescence line, both of which are a function of the
ionization state and elemental abundances of the reflecting gas
(e.g. Lightman \& White 1988; George \& Fabian 1991; Matt, Perola \&
Piro 1991). Such models have been applied to magnetic CV systems
(Beardmore et al 1995; Done, Osborne and Beardmore 1995; Cropper,
Ramsey and Wu 1996; van Teeseling, Kaastra and Heise 1996), but have
not yet been extended to the non--magnetic systems.

A further possible distortion of the hard X--ray spectrum in DN is
from complex absorption. P Cygni line profiles are seen in these
systems in the UV SiIV, CIV and NV resonance lines, but only in
outburst, when the accretion rate is high (see e.g. the recent review
by Drew 1993). The mass loss rate and ionization structure in this
wind is poorly known, but the models which have most success in
matching the blueshifted UV absorption features are optically thick at
soft X--ray energies (e.g. Mauche and Raymond 1987; Hoare and Drew
1991; 1993).  Thus partially ionized absorption may play an important
role in determining the observed soft X--ray spectrum in outbursting DN
spectra, though the wind declines rapidly after the outburst peak
(Verbunt et al 1984), so this is unlikely to be important for the 
quiescent emission. 

These spectral distortions will bias the results from simple thermal
model fits to the data. This is important as recent exciting
developments in accretion disk structure have attempted to include a
self consistent description of the boundary layer emission. In the
quiescent state the disk density is low, so local cooling may be 
inefficient enough that {\it radial} energy transport 
is also important.  Numerical solutions by Narayan and Popham (1993)
including these advected energy terms give predictions
of the temperature--emissivity structure of the boundary layer. 
However, as both
reflection and ionized absorption give an observed spectrum that is
flatter than the intrinsic continuum, their neglect can give
an overestimate of the amount of high temperature material.

SS Cyg is the brightest DN at in hard X--rays, so is the obvious
candidate for detailed spectral studies. Here we analyse the available
archival data from GINGA (2 spectra) and ASCA (1 spectrum). These show
the system in three different states, quiescence, decline from a
normal (fast rise) outburst, and in anomalous (slow rise) outburst,
respectively. We evaluate the importance of reflection and partially
ionized absorption, and so obtain the underlying intrinsic continuum
temperature--emissivity distribution of the hot plasma. These data
have been previously reported by Yoshida et al (1992), Ponman et al
(1995) and Nousek et al (1994) but these authors only fit the spectra
with simple thermal models. Here we include reflection and ionized
absorption, and explicitally fit continuous temperature distribution
models to the underlying intrinsic spectral shape. 

We clearly detect the signatures of reflection in all the spectra.  In
outburst this is consistent with the hard X--ray emitting plasma
forming a corona over the white dwarf, forming a reflection spectrum
from both the white dwarf surface and inner disk. However, in
quiescence there is significantly less reflection, which we interpret
as indicating that the inner disk is disrupted or optically thin
during quiescence. The ASCA data also show partially ionized absorption,
as expected from the outflowing wind. 

The abundances of the elements can be constrained, especially in the
ASCA data. The line emission is significantly weaker than expected
from solar abundances. We examine possible deviations from coronal
equilibrium, but conclude that none of these can cause the observed
line deficit. The high iron abundance relative to the other elements
in the Anders \& Grevasse (1989) 'solar' compilation is inconsistent
with the data. Newer 'solar' abundance determinations (Grevasse,
Noels \& Sauval 1996) give a relative iron abundance much closer to the
Morrison \& McCammon (1983) values, which are similar to the elemental
abundance ratios in the ASCA data presented here.

The underlying continuum is well described by a continuous (power law)
temperature--emissivity distribution of hot material.  In quiescence
this is strongly weighted towards high temperatures, so that it is
indistinguishable from a single temperature component.  This is
inconsistent with the expected cooling of hot gas. Either there is
some reheating process or the cool components are masked by partial
absorption, perhaps by an inner disk which is optically thin to
electron scattering which photoelectrically absorbs the soft X--ray
emission from the boundary layer below the disk. By contrast, the
outburst spectrum is dominated by cool components, and cannot be
described by a single temperature plasma. While the emission is
significantly softer, the total luminosity is similar to that in
quiescence, assuming that the material cools down onto the
photosphere.

\section{THEORETICAL MODELS}

\subsection{Continuum Emission}

If gas is heated up to some temperature $T_{max}$ and then allowed to
cool down to the photospheric temperature $T_{bb}$ at constant
pressure in a constant gravitational potential so that the luminosity
is solely driven by the temperature drop then the resulting spectrum
can be described by a sum of thermal spectra from $T_{max}$ to $T_{bb}$
such that 
$$L(\nu)\propto \int_{T_{bb}}^{T_{max}} {\epsilon(T,n^2,\nu)\over \epsilon(T,n^2)}
dT$$
where $\epsilon(T,n^2,\nu)$ is the bremsstrahlung spectral emissivity
and $\epsilon(T,n^2)$ is the total (integrated over $\nu$)
bremsstrahlung emissivity. 
More general {\it cooling flow} models
include a weighting factor factor $(T/T_{max})^\alpha$ in the
integrand (see e.g. Johnson et al 1992; Done et al 1995). 

Explicit calculations of the emission from the boundary layer are
extremely difficult due to the strongly shearing and probably
turbulent nature of the flow.  Any accretion disk problem is
inherently two dimensional (at least) as the matter flow is radial
while the energy is transported vertically (through radiation) as well
as radially (through advection).  Reducing the problem to 
1 dimension by vertical averaging of the
accretion disk structure (the standard Shakura--Sunyaev approach)
makes the equations far more tractable. This is the approach used by
Narayan \& Popham (1993) and Popham \& Narayan (1995) to calculate
the temperature structure of the boundary layer. Their models then 
give the predicted temperature--emissivity distribution in
the boundary layer.
Figures 1a and b show spectra calculated from the models of Narayan \& Popham 
(1993) for the quiescent emission
for a white dwarf of mass $M=M_\odot$ and radius $R=5\times
10^8$ cm for a mass accretion rate of $\dot{M}=10^{-9.5}$ and
$10^{-10.5}$ $M_\odot$ yr$^{-1}$, respectively. The dotted lines
in these figures show a comparison with an $\alpha=0$ cooling flow model
of maximum temperature equal to that of the highest temperature
emission present in the advective boundary layer model,
which is $8.78$ and $16.4$ keV, respectively. The two models are
remarkably similar at low $\dot{M}$, but begin to diverge at higher 
mass accretion rates. In outburst however, their models give no
hot optically thin component to the emission (Popham \& Narayan 1995),
probably because of the limitations of the 1 dimensional approach
whereas the hard X--ray emission is thought to be from material 
escaping from the boundary layer to form a corona over the
white dwarf/disk.

\begin{figure}
\plotone{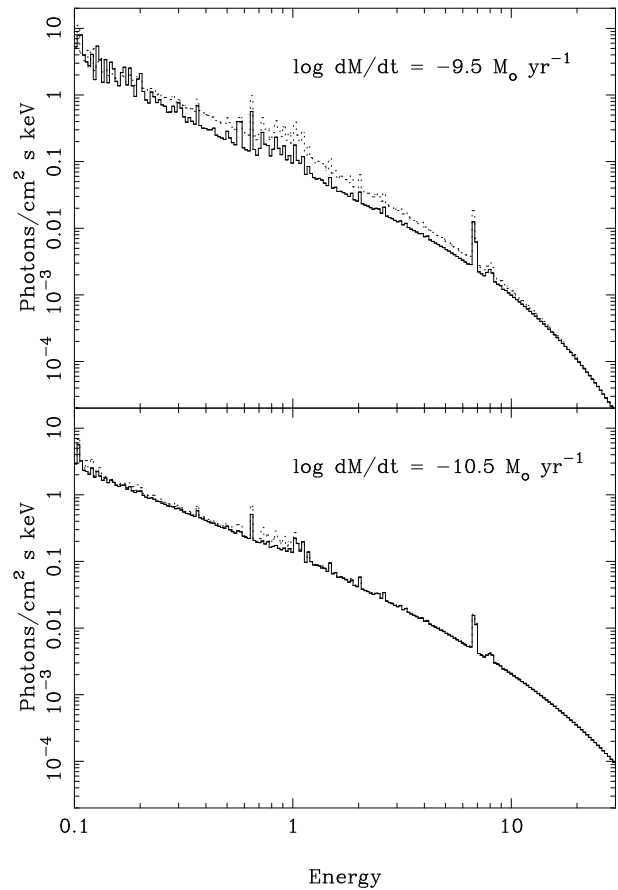}
\caption{Spectra calculated from the models of Narayan \& Popham (1993) for the quiescent
emission from a $1M_\odot$ object accreting at $\log \dot{M}=-9.5$
(a: top panel) and $-10.5 M_\odot$ yr$^{-1}$ (b: lower panel) are shown by
the solid lines. The dotted lines give the comparison spectrum
expected from plasma of solar abundance heated to the same maximum temperature as in the
Narayan \& Popham (1993) models, cooling by radiation alone, at
constant pressure in a constant gravitational potential. The models
are very similar, especially at low mass accretion rates.}
\end{figure}

Calculations in (at least) 2 dimensions are required in order to
follow coronal structure.  Kley (1989) and Hujeirat (1995) solve
numerically a fully 2 dimensional set of (simplified) equations
describing the hydrodynamic and thermal properties of the gas. The
results of Kley (1989) are especially relevant as these include most
of the white dwarf surface in the calculation, and so can follow any
hot, optically thin gas which expands to form a corona around the
white dwarf, while the grid used by Hujeirat (1995) only covers the
equatorial belt of the white dwarf. Kley (1989) does indeed show that
an optically thin, hot ($\sim 6$ keV for his assumed system
parameters) corona forms around the white dwarf and inner disk during
outburst. The temperature and density contours can in principle 
be combined to give the temperature--emissivity distribution 
(which is the observed quantity) but in practice the published data 
plots are too complex to convert into this form so we cannot directly
compare these calculations with the data. 

However, even a two dimensional approach is probably not adequate, as
this cannot describe oblique shocks in the plane of the disk which are
perhaps the most likely outcome of the disk/star interaction. Even
more seriously, all these models (both 1 and 2 dimensional) assume
that the viscosity in the boundary layer and the disk is produced the
same mechanism (generally assumed to be turbulent convective motions) and so use the same
parameterisation to describe the viscosity in both regions. This is at
odds with recent results which show that convective turbulence
transports angular momentum {\it inwards} not outwards as is required
for accretion (Balbus, Hawley \& Stone 1996), and that the magnetic
dynamo models (Tout \& Pringle 1992) based on the magnetohydrodynamic
instabilities (Balbus \& Hawley 1991) are the most likely source of the
strong viscosity inferred for accretion disks. The viscosity in the
boundary layer may then be very different from that in the disk since
it is not at all obvious that the MHD dynamo can work outside of the
ordered velocity field in the disk (since it uses Keplarian shear
velocities to transform radial magnetic field to azimuthal, the Parker
instability to transform azimuthal field into vertical, and the
Balbus--Hawley instability to regenerate radial field from
vertical. The dissipation required for viscosity then comes from
magnetic field reconnection from opposing vertical field lines:
Armitage, Livio \& Pringle 1996). 


\subsection{Reflection from the White Dwarf Surface}

The intrinsic continuum can be reflected by any optically thick
material.  The reflection models calculated to date have generally
assumed the specific geometry of a (pointlike or extended) source
illuminating a flat slab (Lightman and White
1988; Matt, Perola and Piro 1991; George and Fabian 1991; Zycki et al
1994; Magdiarziz and Zdziarski 1995; Van Teesling et al 1996).  Here
the geometry of the boundary layer emission is more complex, as
we expect that the X--ray emission will be produced in an equatorial
belt around the white dwarf. The vertical extent of this belt is model
dependent.  The advective boundary layer models of Narayan and Popham (1993) have 
$h/r\sim 0.3-1$ 
(R. Popham, private communication), while King and
Shaviv (1984) and Pringle and Savonije (1979) envisage the quiescent
boundary layer as extending as a corona over the whole of the white
dwarf surface. In outburst all the models agree that the optically
thick boundary layer should be extremely compact both in height and in
radial extent. However, the structure of the residual optically thin
hard X--ray emission that can be seen in outburst is less clear, and
such high temperature gas is again likely to expand adiabatically out
of the equatorial plane before it cools (Pringle and Savonije 1979;
King and Shaviv 1984).  Thus we use the vertical extent of the
boundary layer as a free parameter.

The radial extent of the boundary layer is observed to be very small,
$\le 1.15 R_{WD}$, from eclipse mapping of the X--ray emission in the
quiescent dwarf novae HT Cas (Mukai et al 1996). Interestingly, the
width of the radiative boundary layer in the models of Narayan and
Popham (1993) is $1.15R_{WD}$ for mass accretion rates of $\dot{M}=
10^{-9.5} M_\odot$ yr$^{-1}$, and {\it increases} for smaller mass
accretion rates. This is in potential conflict with the data as
the hard X--ray flux from HT Cas only requires an
accretion rate of $\sim 2\times 10^{-12} M_\odot$ yr$^{-1}$ (Mukai et
al 1996), but the fractional contribution of the hard X--rays to the
bolometric luminosity is highly uncertain and could be very small (Van
Teeseling, Beuermann and Verbunt 1996).
Here we follow the observations and assume that the boundary layer is
thin in the radial direction.

The boundary layer emission will clearly illuminate the white dwarf
surface, but the extent to which this can be seen depends on the
accretion disk geometry.  In outburst it is generally supposed that
the disk is optically thick and has an inner radius which is very
close to the white dwarf surface (e.g. Popham and Narayan 1995).
However, there is considerable debate on the structure of the
accretion disk during quiescence.  Several models have suggested that
there is no inner disk in quiescence, in order to explain the delay in
the onset of the UV flux rise during outburst with respect to the
optical (as seen by e.g. Mauche et al 1995).  Meyer and
Meyer--Hofmeister (1994) proposed that the inner disk could be
evaporated by a coronal siphon flow; Livio and Pringle (1992) proposed
that the inner disk formation during periods of low mass accretion
rates was inhibited by low level ($B\sim 10^4-10^5$ G) magnetic
fields, while Armitage et al. (1996) proposed that the inner disk is
depleted by a magnetic wind.  Alternatively, the inner disk may be
present in quiescence, but not be optically thick. This is suggested
by Mukai et al (1996) to explain the marginally narrower X--ray than
optical eclipse ingress/egress width in HT Cas. If the accretion disk
has a column of $\sim 10^{23-24}$ cm$^{-2}$ then it is optically thick
to photoelectric absorption of the 2--3 keV flux (which dominates the
ASCA countrate) emitted by the boundary layer below the disk but is
optically thin to the optical emission from the white dwarf
photosphere as long as the disk density is less than $\sim 10^{15-16}$
cm$^{-3}$ so that the free--free opacity is small. 
Thus we assume that an
optically thick disk extends down to the boundary layer in outburst,
but that in quiescence the inner disk {\it may} be disrupted or optically thin.

The proposed geometries are shown in Figure 2a and b.  The boundary
layer half thickness subtends an angle $\beta$ at the center of the
white dwarf, and the system is viewed at some inclination angle $i$
(measured from the normal to the disk plane). The appendix shows that
the reflection component from the white dwarf surface in SS Cyg, where
$i=37^\circ$ (Cowley et al 1980), can be approximated by that from an
isotropically illuminated slab viewed at $60^\circ$, irrespective of
$\beta$. 

\begin{figure}
\plottwo{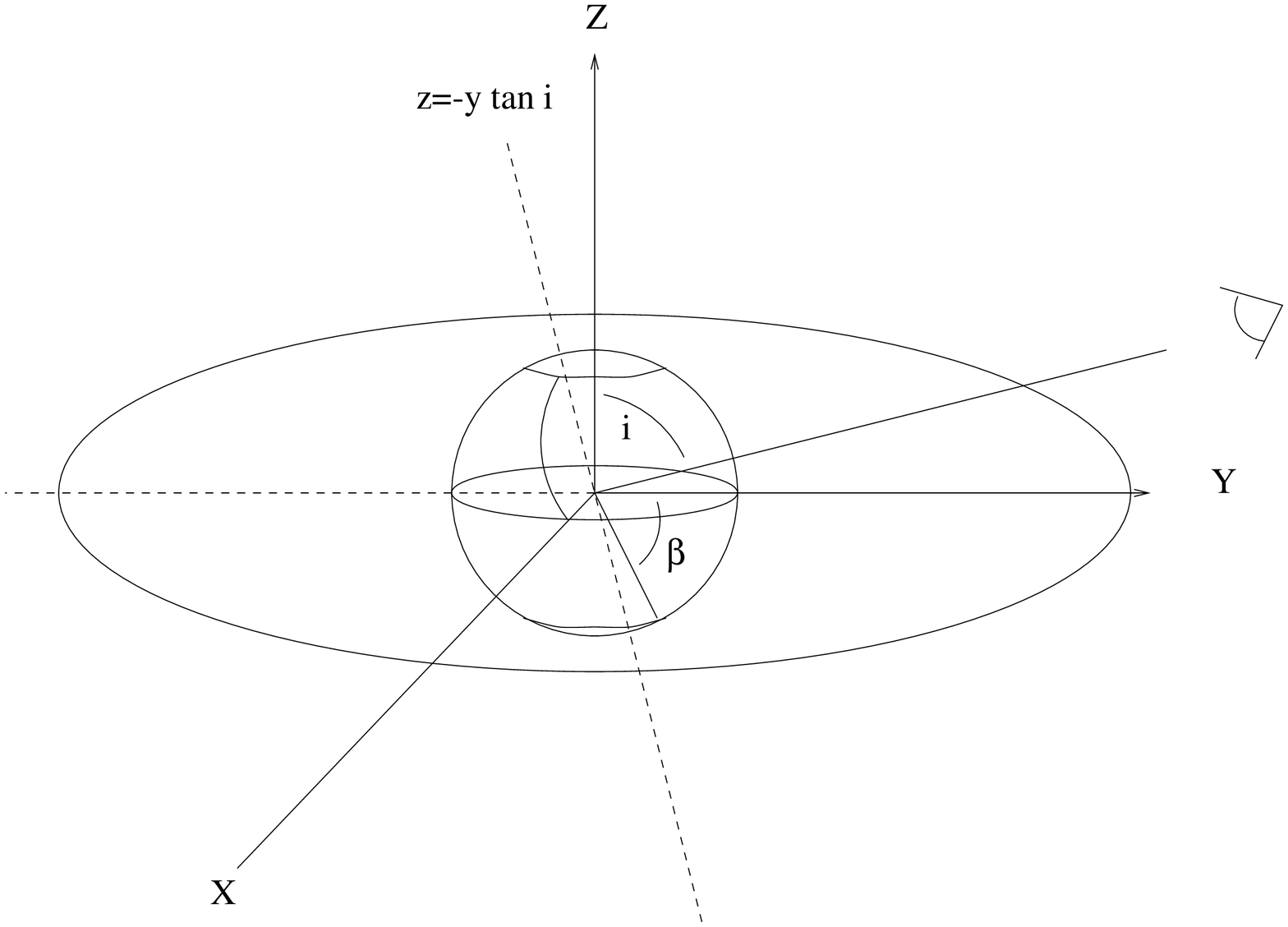}{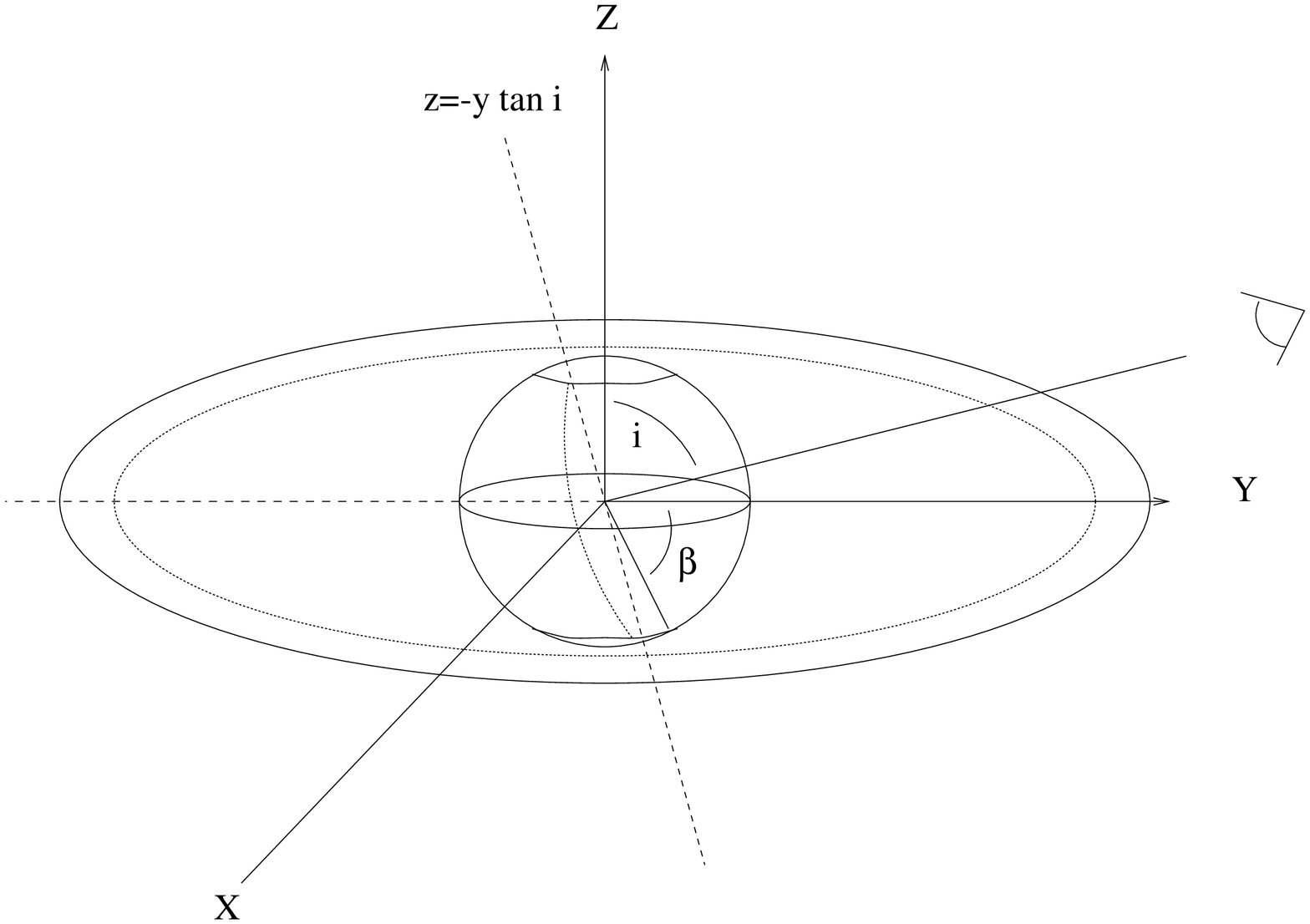}
\caption{The assumed geometry of the boundary layer. The X--ray emission 
forms a corona over the white dwarf surface, extending up to an angle
$\beta$ towards the pole. The observer is assumed to be on the Y axis,
viewing the system at some inclination $i$ with respect to the normal to 
the accretion disk in the X-Y plane. In 'a' this disk is assumed to
be optically thick all the way down to the boundary layer, whereas in 'b'
the inner disk is either not present or is optically thin so that the
emission from the boundary layer below the equatorial plane can also
be seen}
\end{figure}

\subsection{Reflection from the Disk}

There should be an additional reflection spectrum from
the optically thick disk if it does extend down to the inner boundary
layer (see figure 2a). This is not so simple to quantify, firstly as
the geometry is complex (the reflection spectrum from the region of
the disk shadowed by the white dwarf cannot be seen by the observer)
and secondly because the amount of reflection depends also on the
distribution of incident photon angles, and is larger for large
incident angles such as are predominantly expected here. Photons
which skim along the surface of the disk have a large path length
through the material. They are are then absorbed/reflected in the
upper layers of the disk, so it is very easy for any reflected photons
to escape (e.g. George and Fabian 1991). There is also the 
problem that the reflector intercepts not just the intrinsic spectrum but also
the reflection spectrum from the white dwarf surface. Thus we will just treat the
limiting cases of an infinitely thin disk illuminated by a boundary layer with 
$\beta=0$ and $\beta=\pi/2$.
The former case is
similar to that of a point source above a disk, as for each point on
the boundary layer near the equator the emission is limited to a
hemisphere pointing away from the white dwarf surface, and half of
this will intercept the disk while the other half will not. 
Thus the effective solid angle subtended by the disk is $\Omega/2\pi=1$.
In the
latter case, where the emission extends all the way up to the poles,
the reflection spectrum normalisation is smaller. Radiation emitted at the
poles is again confined to a hemisphere pointing away from the white
dwarf surface but this does not intersect the disk at all, and so
contributes only to the continuum emission rather than to the disk
reflection. In fact the solid angle subtended by the disk to an optically
thick X--ray emitting sphere is 
$\Omega/2\pi=1/2$ (Chen and Halpern 1989), so as $\beta\to \pi/2$
the additional 
reflection component is then 
$0.5\times$ that expected from an isotropically illuminated
disk viewed at inclination $i$. For SS Cyg the angles are such that
an $\Omega/2\pi=1/2$ disk reflection component viewed at $37^\circ$ is 
almost identical (in shape and normalisation) to that from
the white dwarf surface ($\Omega/2\pi=1$) viewed at the mean angle of $60^\circ$,
so the additional reflection spectrum expected from the inner disk (if it exists) 
is $1-2\times$ that expected from a slab viewed at $60^\circ$.

\section{OBSERVATIONS}

SS Cyg was observed twice with GINGA (2--20 keV), once in quiescence (25--27 Nov
1987) and once during decline of a normal outburst (2--5 Dec
1990). The data were extracted in 64 second binning using standard
selection criteria (maximum pointing offset from the source of
$0.4^\circ$, soft electron rates of SOL $\le 10$, magnetic rigidity between 
$7\le $ COR$\le 20$, surplus count rate above the upper
discriminator of SUD$\le 10$, and Earth elevation angle between $6\le$ YELEV$\le 120$). 

The GINGA proportional counters are split into two separate layers,
top and mid, where the mid layer is more sensitive above 15 keV, but
has very little effective area below 8 keV (Turner et al 1989). The
data mode chosen for the majority of both SS Cyg observations was
MPC2, in which the top and mid layer are combined, rather than the
usual faint source data mode of MPC1 where the mid layer spectra can be
discarded to increase signal--to--noise.  The standard 'universal'
background subtraction technique, in which blank sky observations over a
period of several months around the date of the observation are used
to calculate the contribution from the diffuse cosmic X--ray
background, particle background and the induced satellite radioactive
decay components (Hayashida et al 1989; Williams et al 1992) is
optimised for MPC1 mode data, so is not necessarily appropriate
here. Also, SS Cyg is in the galactic plane ($b=-7$) so the spectrum
at the lowest energies can also be contaminated by the diffuse
galactic background.  Instead we use a 'local' background (Williams et
al 1992), where data from a nearby (in both time and galactic
coordinates) were also extracted, using the same selection criteria as
the source except that a larger range in geomagnetic parameters
(SUD$\le 14$) and offset angle ($\le 0.5^\circ$) is allowed. This
blank sky data is then used to model the background components in the
source data.  For the 1987 data the background observation is not
sufficiently long to sample all the same background levels as seen
during the source observation so we exclude parts of the SS Cyg
observation with SUD$\ge 8$. Two nearby backgrounds were used to 
model the 1990 data as the background taken between the two main blocks
of source data was noisy. Systematic errors induced by using only a short
segment of background data are included by increasing the error bar
on the background by a factor 1.5. 
This is a conservative procedure and probably results in 
the overly small values of $\chi^2$ derived in Sections 4.1 and 4.2. 
The background subtracted data is then
attitude corrected in order to produce the final dataset, and a
systematic error of 0.5 per cent is added to take account of residual
uncertainties in the determination of the instrument response.

The default (Rev 1) screened ASCA (0.6--10 keV) data for each instrument were
taken from the archive. These were extracted with elevation angle $\ge
10^\circ$, geomagnetic rigidity $\ge 6$ GeV/c, pointing within
$0.01^\circ$ of the target position, not in the South Atlantic Anomaly
or in fine guidance mode. For the SIS data there is the further 
criteria that the bright Earth angle is $\ge 20^\circ$, and 
all faint mode data are converted to bright mode, and 
combined with the original bright mode data. Source spectra for
the SIS and GIS were extracted over a circle of radius 4 and 6
arcminutes, respectively, and background was taken from the same image
as the source data. The effective area curve for each instrument was
derived from the source position data and this, together with the the
standard response matrices (s0c1g0234p40e1\_512v0\_8i.rmf,
s1c3g0234p40e1\_512v0\_8i.rmf, gis2v4\_0.rmf, gis3v4\_0.rmf) give the
instrument response. The source data were then grouped to a minimum 
of 20 counts per bin so that $\chi^2$ fitting is appropriate. 

\section{SPECTRAL FITTING}

The data were fitted using the XSPEC spectral fitting package version
9.00 (Arnaud 1996). 
All error bars or upper limits quoted are $\Delta\chi^2=2.7$ (90
\% confidence level for 1 free parameter). We caution that some parameters are
strongly coupled (especially in the low spectral resolution GINGA data),
so the allowed strengths of e.g. the Gaussian line and reflection continuum
are strongly (anti)correlated. 
The equivalent widths of any additional Gaussian
line components are quoted by re--evaluating the model on a dummy
response matrix with 1000 energy grid points between 0.2 and 20
keV. This means that the 6.4 keV line is calculated with respect to
the line--free continuum at 6.4 keV, whereas with the instrumental
energy binning the 'continuum' can be contaminated by the hot plasma
line emission.  The interstellar column density to SS Cyg is fixed at
$N_H=3\times 10^{19}$ cm$^{-2}$, as determined from the depth of
narrow UV absorption lines seen against the bright outburst spectrum
(Mauche, Raymond and Cordova 1988).

For single temperature hot plasma spectra we use the {\tt mekal} model
in XSPEC (see the XSPEC users manual for a complete description of the 
supported models), 
while for a power law emissivity--temperature distribution plasma
we use a modification of the {\tt cemekl} code in XSPEC, 
where the luminosity of each temperature component is weighted by a
factor $(T/T_{\rm max})^\alpha$.  The code as released
calculates the integral over temperature logarithmically, without
modifying the step size from $dT$ to $d \log T$.  Thus the $\alpha$
produced by the code is 1 larger than the true weighting factor.  In
all that follows the $\alpha$ quoted is the XSPEC weighting
factor.  Numerically the code also had to be modified by changing the
order of integration (starting with the highest temperature component
rather than the lowest) so that the spectrum changes smoothly with
changing $T_{\rm max}$. The density in both plasma codes is fixed at $n=1$ cm$^{-3}$.
Reflection of the continuum spectrum is calculated from the code
of Magdziarz and Zdziarski (1995), with the inclination
angle fixed at $60^\circ$ as appropriate for SS Cyg (see section 2), 
but with the fractional contribution left as a free
parameter. All these models assume that the abundances with respect to H
of all the elements
scale together, and that this fractional abundance relative to the Solar 
determination of Anders and Gervasse (1989) is a free parameter.
The iron fluorescence line that should accompany the reflection continuum
is included as a separate narrow Gaussian, with line energy of 6.4 keV 
being appropriate for a mainly neutral reflector. These four
model components (single temperature plasma, power law temperature--emissivity 
distribution plasma, reflection continuum and Gaussian emission line)
are denoted by T, plT, R and g respectively in all the tables. 

There is also the possibility of complex absorption.  In quiescence
the inner disk may be present but be optically thin to electron
scattering (section 2.2). This may give rise to partial absorption
of the X--ray emission. The fraction of the emission that is occulted
by the inner disk can be derived from the difference in the emitting
area observed with and without a disk (i.e. the difference in
normalisation factors in equations 1 and 2 of the appendix).  For SS
Cyg at an inclination of $37^\circ$
this gives rise to a maximum expected covering fraction of 0.5 if
the boundary layer is concentrated in the equatorial plane ($\beta\to
0$) or a minimum covering fraction of 0.2 for $\beta\to 90^\circ$.  We
assume that the optically thin disk is neutral and model this by the
{\tt pcf} code in XSPEC.  
The partially ionized absorption expected in outburst from
transmission of the X--ray spectrum through the outflowing wind is
modeled by the {\tt absori} code in XSPEC. The ion populations are
found by balancing the temperature dependent dielectronic and
radiative recombination rates against photoionization as in Done et al
(1992). The thermal balance equations are not calculated self
consistently. Instead the temperature of the material is fixed at
$5\times 10^4$ K, as appropriate for the derived low ionization states
(see section 4.3 and e.g. Kallman and McCray 1982 figures 21 and 22).
The code assumes that the ionization balance is determined by a power law
X--ray spectrum.
The continuum from SS Cyg can be roughly
characterised by a power law of photon index $\Gamma=2$ in outburst
and $\Gamma=1.8$ in quiescence, so we fix the index at the relevant
value for each spectrum. Model calculations of the ionization state of
the wind have tended to concentrate on photoionization by the putative
boundary layer blackbody emission but the harder X--ray emission will
also contribute to the ionization balance. This especially important
in determining the abundances of the ions which absorb in the X--ray
bandpass, such as OVII.  Both absorption models use the fixed abundances
of Morrison \& McCammon (1983) and are denoted by pcf and pia
respectively in all the tables.

\subsection{GINGA 2--20 keV Spectrum (1987): Quiescence}

Results from all the spectral fits in this section are  detailed in 
table 1. Since these data are the same as those used by Yoshida et al
(1992) we first check for consistency by using their model of a 
bremsstrahlung spectrum with narrow Gaussian emission line. 
This gives a good fit to the data ($\chi^2_\nu=15.3/22$), with 
derived continuum and line parameters which are consistent
with those quoted by Yoshida et al (1992). 
However, a bremsstrahlung spectrum
is physically unlikely as there will be elements other than just pure 
hydrogen in the accreting material. Thus we replace this {\it ad hoc}
continuum and line with a hot plasma spectrum in which the line
emission is calculated self consistently, and fix the absorption at the 
interstellar value. This gives a poor fit to the spectrum,
with $\chi^2_\nu=39.6/24$ for a temperature of $15.2\pm 0.7$ keV and
abundance of $0.71\pm 0.10$. Including a second temperature component,
as has often been suggested (Yoshida et al 1992; Nousek et al 1994)
does not give a significantly better fit ($\chi^2_\nu=32.6/22$ for
$kT_1=9.2^{+2.9}_{-3.0}$ keV and $kT_2=57^{+43}_{-37}$ keV).  Adding
further separate temperature components does not result in any
significant improvement in the fit.

Such {\it ad hoc} multi--temperature models can approximate the
spectrum of a continuous temperature distribution.  A power law
temperature--emissivity distribution hot plasma model
gives a fit which is not
significantly different ($\chi^2_\nu=37.6/23$) from the single
temperature description, with the data requiring that the most weight
is given to the higher temperature components ($\alpha\ge 1.6$).

As the theoretical continuum spectral shape is not known {\it a
priori} we are justified in 
trying an arbitrary emissivity/temperature relation.  
One such model describes the differential emission
measure, $\phi(T)$, as an exponential function of an $k$th order
Chebyshev polynomial, $P_k(T)$, such that $\phi(T)\propto exp(\Sigma
a_kP_k(T)$ (Lemen et al. 1989), where the exponential function ensures
positivity. We modified the {\tt cp6mkl} model in XSPEC so that the
maximum temperature was a free parameter rather than being fixed at 10
keV, redefined the Chebyshev polynomial coefficients to a more standard
form (e.g. Press et al 1988), and
again reversed the order of integration.  Fitting with this
model resulted in a $\chi^2_\nu=33.0/18$, similar to that from the two
temperature plasma description. The derived
emissivity distribution has two peaks at similar temperatures to
those found in the two temperature model so we do not tabulate the
results of this fit. 

\begin{table*}
\begin{minipage}{180mm}
\caption{Quiescent (GINGA) data}
\label{qu}
\begin{tabular}{cccccccccc}                \hline

model & kT\footnote{This is the maximum temperature in the plT models} (keV) 
& A & N\footnote{The normalisation of the plasma model in units of 
$10^{-14}/(4\pi D^2)\times$ Emission Measure} & $\alpha$ & $R$ & EW (eV) 
& $N_H$\footnote{Units of $10^{21}$ cm$^{-2}$} & $\xi$\footnote{Defined as 
$L/nr^2$ where $r$ is distance from material of density $n$ illuminated 
by a source of luminosity $L$} & $\chi^2_\nu$\\
\hline

1T & $15.2\pm 0.7$ & $0.71\pm 0.10$ & $4.1\times 10^{-2}$ & & & & && $39.6/24$\\
2T & $9.2^{+2.9}_{-3.0}$ & $0.69^{+0.14}_{-0.09}$ & $2.7\times 10^{-2}$ & 
& & &&& $32.6/22$\\
   & $57^{+43}_{-37}$ & & $1.7\times 10^{-2}$ \\
plT & $19.7^{+5.7}_{-4.8}$ & $0.74\pm 0.11$ & $0.20$ & $2.8^{+\infty}_{-1.2}$ 
& & & &&$37.6/23$ \\
1T+g & $15.6\pm 0.7$ & $0.42\pm 0.13$ & $4.3\times 10^{-2}$ & & & $170\pm 60$ 
&&& 15.0/23 \\
1T+R+g & $13.0^{+3.0}_{-2.3}$ & $0.35^{+0.15}_{-0.10}$ & $4.2\times 10^{-2}$ &
& $0.5^{+0.7}_{-0.5}$ &$125^{+95}_{-90}$ &&& $13.2/22$ \\
plT+R+g & $15.4^{+8.8}_{-5.4}$ & $0.35^{+0.16}_{-0.11}$ & $0.21$ &
$2.9^{+\infty}_{-1.4}$ & $0.6^{+0.9}_{-0.6}$ & $100_{-85}^{+90}$ &&& $12.2/21$ \\
pcf(plT+R+g)\footnote{Covering fraction fixed at 0.2} & $21_{-5.7}^{+11}$ &
$0.38_{-0.11}^{+0.16}$ & $0.10$ & $1$\footnote{Parameter fixed} & 
$0.6_{-0.6}^{+1.0}$ & $85_{-85}^{+100}$ 
& $16_{-7}^{+17}$ & $0$$^e$ & $10.2/21$ \\ 
pia(plT+R+g) & $21_{-5.8}^{+11}$ & $0.38^{+0.17}_{-0.11}$ & $0.10$ &
$1$$^f$ & $0.6^{+1.1}_{-0.6}$ & $85_{-85}^{+90}$ & $4.7^{+36}_{-3.5}$ & 
$18_{-18}^{+1000}$ & $10.3/20$ \\
\end{tabular}
\end{minipage}
\end{table*}

All the above plasma models give systematic residuals at the iron line
and edge energy. These are shown in figure 3 for the best fitting two
temperature plasma model, and are very suggestive of the presence of a
reflection continuum and its associated iron K$\alpha$ line emission.
Including an additional cold iron line to even the single temperature
plasma model gives a highly significant reduction in $\chi^2_\nu$ to
$15.0/23$.  This line should also be accompanied by a reflection
continuum, but the strength and shape of this continuum is dependent
on the elemental abundances and ionization state of the reflector
as well as the
geometry of the source with respect to the reflector and the
inclination of the reflector with respect to the observer (see Section 2).
We fix the inclination at $60^\circ$, and constrain the abundances in the
reflector to be equal to those of the hot plasma.
This gives only a marginal improvement
in the fit, with $\chi^2_\nu=13.2/22$ compared to the 6.4 keV line fit
described above, but physically it gives a coherent picture of the
source as the amount of 6.4 keV fluorescent line is then consistent
with a reflection origin. 

\begin{figure}
\plotone{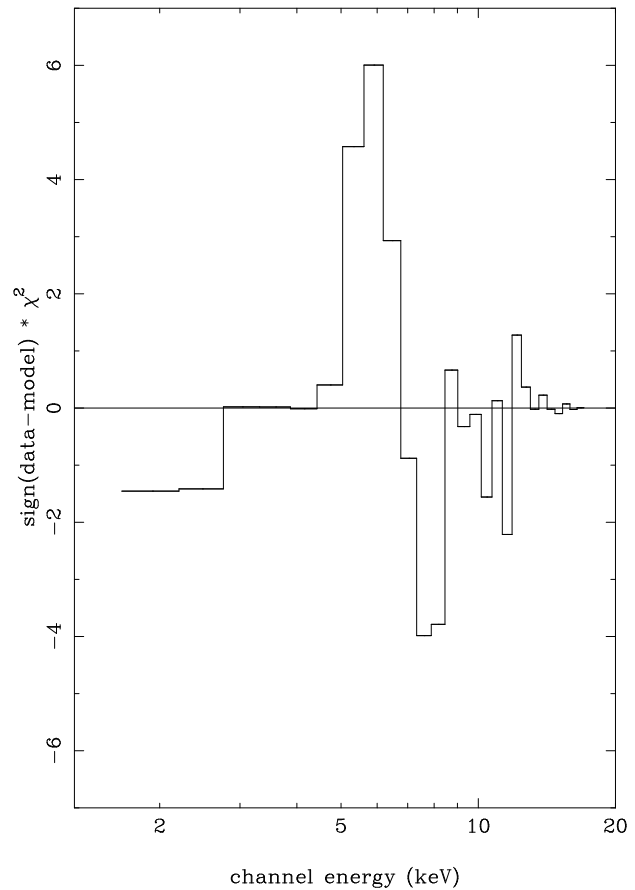}
\caption{The $\chi^2$  residuals of the GINGA quiescent spectrum to a 
two temperature plasma model. There is an excess at $\sim 6.4$ keV,
together with a decrement at 7-8 keV. These features are consistent
with the iron fluorescent line and edge expected from reflection in cold
material.
}
\end{figure}

Replacing the single temperature continuum model with a power law
temperature--emissivity distribution and fitting this, together with
its reflection spectrum and a cold iron line as above, gives no
significant improvement in the fit ($\chi^2_\nu=12.2/21$), with the
incident spectrum being weighted towards the high temperature
components ($\alpha\ge 1.5$).  The total intrinsic 25eV--100 keV
flux in this model is $1.2\times 10^{-10}$ ergs cm$^{-2}$ s$^{-1}$, which  
gives a bolometric luminosity of the optically thin gas of $1.4\times
10^{32}$ ergs s$^{-1}$ assuming a distance of 100 pc (Bailey 1981).  
This implies a
{\it minimum} mass accretion rate of $\dot{M}=10^{-10.4} M_\odot$ yr$^{-1}$ assuming
that the mass of the white dwarf is $0.8 M_\odot$ (Cowley et al 1980)
and its radius is $R_{WD}=7\times 10^8$ cm (Nauenberg 1972). 
The temperature--emissivity distribution derived from the data is
inconsistent with the $\alpha=1$ gas cooling model and the advective boundary layer
model at these accretion rates (see Figure 1), while freezing
$\alpha$ at unity gives significant excess emission at low energies.
The mass accretion rate given above may be underestimated by its neglect of possible
optically thick emission from the boundary layer, but such corrections
are likely to be small as the optically thick emission in quiescence
seems to be weak (Ponman et al 1995). 
One possible way for the spectra to be made consistent with these
models is to add in excess absorption. The data clearly are already
well fit statistically but the addition of partial covering by cold
material with the absorbed fraction fixed at 0.2 (see Section 3) gives
an excellent fit ($\chi^2_\nu=10.2/20$) for a column of $N_H\sim
2\times 10^{22}$ cm$^{-2}$ with the derived temperature--emissivity
relation fixed at $\alpha=1$.

Alternatively, the excess absorption can be described by complete covering by 
partially ionized material (the {\tt absori} model
in XSPEC with the power law index fixed at $1.8$). Again, this gives
a good fit ($\chi^2_\nu=10.3/20$) with the $\alpha=1$ continuum model
for a column of $N_H\sim 5\times 10^{21}$ cm$^{-2}$. Both absorption models
increase the intrinsic (absorption corrected) 
25eV--100 keV flux to $1.4\times 10^{10}$ ergs cm$^{-2}$ s$^{-1}$. 

\subsection{GINGA 2--20 keV Spectrum (1990): Decline from Normal Outburst}

These data were reported by Ponman et al (1995), who noted that the
spectrum is significantly fainter and softer than that seen in
quiescence.  They split the observation into two sections in order to
compare them with simultaneous ROSAT survey data. However, they noted
that there was no significant change in hard X--ray spectral shape
between the two segments, even though the softer ROSAT data, which are
dominated by a separate low temperature component, show substantial
variability.  Thus here we co--add all the data to get maximum signal
to noise. A single temperature bremsstrahlung continuum and (broad)
Gaussian emission line gave results consistent with those of Ponman et
al (1995).

We test the same series of models as described above, and details of
all the spectral fits are given in table 2. Again, neither single nor
multiple temperature models give a good fit to the continuum. However,
unlike the quiescent data, adding a 6.4 keV iron emission line to the
single temperature continuum does not produce a statistically
acceptable fit to the data. Similarly, a single temperature continuum
together with a reflection continuum and line fails to fit the data
($\chi^2_\nu=51.1/22$).  A good phenomenological description of the
data can only be achieved by adding a 6.4 keV emission line to an
intrinsically multi--temperature continuum, such as a two temperature
model ($\chi^2_\nu=19.0/21$), or a continuous power law
temperature--emissivity distribution ($\chi^2_\nu=18.4/22$). The lower
value of the temperature weighting index $\alpha$ in this latter fit
indicates the much softer spectrum seen in outburst compared to that
in quiescence.

\begin{table*}
\begin{minipage}{180mm}
\caption{Decline from normal outburst (GINGA) data}
\label{no}
\begin{tabular}{cccccccc}                \hline

model & kT\footnote{This is the maximum temperature in the plT models} (keV) 
& A & N\footnote{The normalisation of the plasma model in units of
$10^{-14}/(4\pi D^2)\times$ Emission Measure}  & $\alpha$ & $R$ & EW (eV) & $\chi^2_\nu$\\
\hline
1T & $6.9\pm 0.3$ & $0.71\pm 0.08$ & $1.34\times 10^{-2}$ & & & & $73.5/24$\\
2T & $4.6\pm 0.6$ & $1.0^{+0.18}_{-0.13}$ & $1.13\times 10^{-2}$ & &&& $46.0/22$\\
   & $100^{+\infty}_{-70}$ & & $3.18\times 10^{-3}$ \\
plT & $12.5^{+2.8}_{-2.1}$ & $0.94\pm 0.13$ & $3.05\times 10^{-2}$ & 
$1.00^{+0.54}_{-0.32}$ & & & $55.1/23$\\
1T+g & $7.1^{+0.3}_{-0.4}$ & $0.42\pm 0.12$ & $1.39\times 10^{-2}$ & & &
$240^{+110}_{-100}$ & $52.6/23$\\
1T+R+g & $6.4^{+0.9}_{-0.7}$ & $0.41\pm 0.11$ & $1.39\times 10^{-2}$ &
& $0.5^{+0.8}_{-0.5}$ & $180^{+130}_{-105}$ & $51.1/22$\\
2T+g & $1.15^{+2.0}_{-0.64}$ & $0.41^{+0.21}_{-0.14}$ & $5.2\times 10^{-3}$ 
& & & $350_{-120}^{+140}$ & $19.0/21$\\
     & $8.6^{+6.6}_{-0.9}$ & & $1.2\times 10^{-2}$ \\
plT+g & $16.4^{+4.7}_{-3.3}$ & $0.50^{+0.16}_{-0.15}$ & $2.21\times 10^{-2}$ & 
$0.51^{+0.35}_{-0.25}$ & & $340^{+110}_{-100}$ & $18.4/22$ \\
plT+R+g & $9.6^{+3.4}_{-1.7}$ & $0.35^{+0.13}_{-0.09}$ & $2.34\times 10^{-2}$ & 
$0.43^{+0.41}_{-0.38}$ & $2.2^{+1.8}_{-1.5}$ & $200^{+130}_{-120}$ & $12.2/21$\\

\end{tabular}
\end{minipage}
\end{table*}

Since the continuous temperature distribution seems more physically
likely as well as giving a better fit to the data we use this as our
intrinsic spectrum and include its corresponding reflection continuum
in the fit as well as the 6.4 keV line emission. This gives a
significant reduction in $\chi^2_\nu$ to $12.2/21$, so that in these
data the reflection continuum itself is detected separately from the
line emission. The model here assumes that the reflecting material is
neutral, so that the reflected edge is at 7.1 keV. This is at first
sight incompatible with the results of Ponman et al (1995), where the
fitted edge energy of 8.3 keV implied highly ionized material.
However, with the limited resolution of GINGA, the derived energy of
spectral features is strongly dependent on the assumed continuum form
(see e.g. Beardmore et al 1995 for a similar example of this effect),

Including a warm absorber (with illuminating power law index fixed at
$\Gamma=2$) does not result in a significant improvement to the fit
($\chi^2_\nu=10.2/19$).  We do not tabulate the results of this fit as
the parameters of the partially ionized column are largely
unconstrained. The intrinsic (absorption corrected) 25eV-100 keV flux
is $6.8\times 10^{-10}$ ergs cm$^{-2}$ s$^{-1}$ and $5.2\times
10^{-11}$ ergs cm$^{-2}$ s$^{-1}$ for models with and without ionized
absorption, respectively.

\subsection{ASCA 0.6--10 keV Spectrum (1990): Anomalous Outburst}

We fit all 4 ASCA datasets simultaneously, and full details of the
derived parameters are given in table 3. The overall shape is 
similar to that of the GINGA decline from outburst data, i.e. much softer 
and fainter than the quiescent spectrum in the 2--10 keV band. 
As shown by Nousek et al
(1994), neither a single temperature plasma nor two plasma components
give a good description of the spectrum, with $\chi^2_\nu=3799/1159$
and $1331/1157$, respectively.  A three temperature model is required
in order to give a statistically acceptable fit
($\chi^2_\nu=1145/1155$). A continuous (power law)
temperature--emission measure distribution gives an even better fit
for fewer free parameters, with $\chi^2_\nu= 1117/1158$.  This result
is contrary to the assertion of Nousek et al (1994) that the lack of
intermediate temperature plasma lines indicates that the emission
measure -- temperature distribution is not monotonic.  The apparent
discrepancy is probably due to their their assumption of solar
abundances which led them to expect large line equivalent widths from
any intermediate temperature components, while the line--to--continuum
ratios in the data are a factor 2 weaker than predicted for solar
abundance models. Their preferred phenomenological model fit to the data is a
three temperature plasma, in which two of the components are assumed
to have solar abundances, while the third has zero abundance (a
bremsstrahlung spectrum).  The latter component acts as a means to
dilute the strong line emission predicted by the other two temperature
components.

There is strongly significant additional low--ionization 
iron line emission. Including a narrow Gaussian emission line at 6.4 keV
in the fit gives a reduction in $\chi^2_\nu$ to $1077/1157$,
with line equivalent width of $\sim 140$ eV. 
However, including the reflection continuum that is expected to accompany
such line emission does {\it not} lead to a decrease in $\chi^2_\nu$.
The best fit reflected fraction is close to zero, with an upper limit of
$R\le 0.7$, too small to produce the observed 6.4 keV emission line.


\begin{table*}
\begin{minipage}{180mm}
\caption{Anomalous Outburst (ASCA) data}
\label{ao}
\begin{tabular}{cccccccccc}                \hline
 
model & kT\footnote{This is the maximum temperature in the plT models} (keV) & A 
& N\footnote{The normalisation of the plasma model in units of
$10^{-14}/(4\pi D^2)\times$ Emission Measure} & $\alpha$ & $R$ & EW (eV) & 
$N_H$\footnote{$\times 10^{21}$ cm$^{-2}$} & $\xi$\footnote{Defined as
$L/nr^2$ where $r$ is distance from material of density $n$ illuminated
by a source of luminosity $L$} & $\chi^2_\nu$\\
\hline
1T & $3.43^{+0.10}_{-0.06}$ & $0.21^{+0.05}_{-0.02}$ & $3.0\times 10^{-2}$ &
& & & & & $3799/1159$\\
2T & $0.67^{+0.03}_{-0.01}$ & $0.51\pm 0.06$ & $3.4\times 10^{-3}$ & & & 
& & & $1331/1157$\\
   & $5.45\pm 0.07$ &        & $2.2\times 10^{-2}$ &\\
3T & $0.56^{+0.04}_{-0.09}$ & $0.44\pm 0.06$ & $2.8\times 10^{-3}$ & & & & 
& & $1145/1155$\\
   & $1.10$ &	     & $3.0\times 10^{-3}$ &\\
   & $6.64$ &        & $2.0\times 10^{-2}$ &\\
plT & $18.2\pm 1.7$ & $0.48^{+0.06}_{-0.05}$ & $3.1\times 10^{-2}$ & 
$0.47\pm 0.05$ & & & & & $1117/1158$\\
plT+g & $17.5^{+2.0}_{-1.5}$ & $0.44^{0.06}_{-0.05}$ & $3.1\times 10^{-2}$ &
$0.45\pm 0.06$ & & $140\pm 40$ & & & $1077/1157$ \\
plT+R+g & $16.9^{+2.1}_{-3.2}$ & $0.44^{+0.06}_{-0.07}$ & 
$3.2\times 10^{-2}$ &
$0.46^{+0.06}_{-0.05}$ & $0.08_{-0.08}^{+0.58}$ & $140\pm 40$ & & & 1077/1156\\
pia(plT+R+g) & $12.6^{+3.6}_{-2.6}$ & $0.32\pm 0.05$ & $2.5\times 10^{-2}$ 
& $0.08\pm 0.11$ & $1.3\pm 0.7$ & $130\pm 40$
& $2.6^{+0.7}_{-0.6}$ & $1.4\pm 0.3$ & $1024/1154$ \\
pia(plT+g) & $19.4_{-2.4}^{+3.4}$ & $0.41\pm 0.06$ & $2.4\times 10^{-2}$ & $0.16^{+0.09}_{-0.11}$ 
& & $160\pm 40$ & $2.1\pm 0.6$ & $1.3^{+0.5}_{-0.3}$ & $1036/1155$\\

\end{tabular}
\end{minipage}
\end{table*}

The low energy bandpass of ASCA means that these data are
also sensitive to details of the absorption column along the line of
sight, and physically some X--ray absorption should be present
from the partially ionized wind seen in SS Cyg in outburst. 
The observed X--ray
continuum spectrum can be described as a power law of photon index
$\Gamma\sim 2$, so we use the the model {\tt absori} in XSPEC, with
the photon index fixed to this value. This leads to a highly 
significant decrease in $\chi^2_\nu$ to $1024/1154$ for
$N_H=2.6\times 10^{21}$ cm$^{-2}$ at an ionization parameter of $\xi=1.4$
for a fixed temperature of $5\times 10^4$ K. The 
underlying continuum form also changes, weighting the power law emissivity--temperature
distribution even more towards the low temperature components, with
$\alpha=0.08$, and the result is that the reflection component is
now significantly detected, with $R=1.3$. Models with the ionized absorption
and 6.4 keV Gaussian line but without the reflection continuum
give $\chi^2=1036/1155$. Thus the reflection continuum is significantly
detected in these data, but only when the intrinsic continuum form is 
properly modeled.

The absorption in this partially ionized material has some effect on
the derived bolometric luminosity of the X--ray emitting gas.  Assuming
that material cools down to the photospheric temperature of $\sim 25$
eV (Ponman et al 1995) gives a total flux from the hot plasma of
$5.2\times 10^{-11}$ ergs cm$^{-2}$ s$^{-1}$ (25 eV-- 100 keV), while
correcting for the ionized absorption increases this to $6.1\times
10^{-11}$ cm$^{-2}$ s$^{-1}$, and further correcting for the
interstellar column gives a flux of $11\times 10^{-11}$ ergs cm$^{-2}$
s$^{-1}$.  This is almost identical to the 
(absorption corrected) hot plasma flux seen in the
GINGA quiescent data integrated over the same band, showing that it 
is not necessarily true that there is less emission from the optically
thin plasma in outburst, only that it is substantially softer. 

The spectral resolution and sensitivity of ASCA make it possible to
investigate the abundances of the elements in the hot plasma emission
separately rather than assuming that they all scale together. 
We replace the plasma emission model with one in which the abundance of each
element is a free parameter (the {\tt cevmkl} model
in XPSEC, modified as described in Section 4).  
The reflection abundance ratios are assumed to be solar,
but the overall elemental abundance is a free parameter. This is fit
separately from the plasma abundances as the plasma line emission may
be distorted from that expected from coronal conditions (see section
5.1).  Allowing the plasma iron
abundance to be free relative to the other elements gives a
significantly better fit ($\chi^2_\nu=999.9/1152$, for $A_{\rm Fe}\sim
0.3$, $A_{\rm rest}\sim 0.5$).  Progressively fitting for the
abundance of Si and  S (which are the next strongest expected lines)
gives $\chi^2_\nu=978.7/1151$ and  $973.2/1150$, for
$A_{\rm Fe}=0.28$, $A_{\rm Si}=0.72$, $A_{\rm S}=0.68$
and $A_{\rm rest}=0.41$ (see table 4). A model in which all the other
observable elements (O, Ne, Na, Mg, Al, Ar, Ca, Ni) are free gives
$\chi^2_\nu=964.2/1143$ (C and N are fixed at solar as their K$\alpha$
lines are below 0.6 keV so their abundance cannot be usefully
constrained).  Thus, from fitting coronal equilibrium models
(together with the reflection continuum and line and ionized absorption), 
the data
{\it require} that the elemental abundances are not solar and that the
abundances ratios are not solar either. The model with fewest 
parameters that
can describe the data (where adding in extra elemental abundance
parameters does not significantly improve the fit), is one in which
all the elements are tied together except for Fe which is a free
parameter, and Si and S, which are tied to each other. This gives
$\chi^2_\nu=973.4/1151$, for $A_{\rm Fe}\sim 0.3$, $A_{\rm Si, S}\sim
0.7$ and $A_{\rm rest}\sim 0.4$.  The spectrum was then truncated
above 1.7 keV, so that the abundance of iron can be measured
separately from the iron L complex. The contribution from reflection
(continuum and line) can then be ignored, and the continuum parameters
$kT_{\rm max}=14.0$, $\alpha=0.15$ and $A_{\rm Si, S}=0.70$ were
frozen at their best fitting value above.  This gives
$\chi^2_\nu=244.8/221$ for $A_{\rm Fe}=0.30$ and $A_{\rm rest}=0.42$,
showing that the iron L complex is consistent with the iron K line
strengths.

The abundances in the reflector are harder to quantify, since they
correlate with the amount of reflection (e.g. George \& Fabian
1991). However, the observed 6.4 keV line emission is also dependent
on the abundances in the reflector and so provides an additional
constraint. This is not straightforward, as the iron line equivalent
width depends on the abundances of {\it all} the elements, not just
iron, as the other elements provide opacity to the escaping iron line
photons (George \& Fabian 1991). The functional dependence is also
made more complicated by the fact that the refected line equivalent
width also depends on the spectral shape of the illuminating
continuum, as for hard spectra there is relatively more luminosity
above the iron K edge that can be absorbed to produce the K$\alpha$
line. However, knowing a line equivalent width from one particular
illuminating spectrum we can scale this by the ratio of the absorped
luminosity above the iron K edge to estimate the line equivalent width
produced from reflection of any other spectral shape (Basko 1978).
The abundance effects can then be estimated from George \& Fabian
1991, who show that for Morrison \& McCammon (1983) abundance ratios
then the line equivalent width scales roughly as $A_{\rm all}^{0.2}$
for $A_{\rm all}\le 1$, and that a relative over-- or under--abundance
of iron then gives a roughly proportional change in the line
equivalent width.

The iron line has an angle averaged equivalent width of 110 eV with
respect to a power law {\it illuminating} continuum of photon index
$\Gamma=1.9$ for Morrison \& McCammon (1983) solar abundances (George
\& Fabian 1991). However, we are using Anders \& Grevasse (1989) 
abundances, in which the iron abundance is $1.5\times$ larger than in
Morrison \& McCammon (1983), while all the other abundances are about
the same. Thus with the Anders \& Grevasse (1989) abundances we
predict a line which is $1.5\times$ larger than this (Reynolds, Fabian
\& Inoue 1996), i.e. an equivalent width of 1.7 keV with respect to the 
reflected continuum. The ASCA spectrum of SS Cyg has best fit spectral
parameters of $kT=14$ keV, $\alpha=0.15$, which has 0.81$\times$ fewer
photons that can be absorbed by the iron K edge than the $\Gamma=1.9$
power law, so predicting a line of 1.4 keV equivalent width with
respect to the reflected continuum. If the abundance ratios in the
reflector scale together then for $A_{all}\sim 0.3$ the line
equivalent width is reduced by a factor $\sim 0.79$, i.e. 1.1 keV.
This is in strong contrast to the best fit model above (where the line
and continuum from reflection are fit separately) in which the iron
line equivalent width with respect to the reflection continuum is
$\sim 0.6$ keV. Fitting a reflection continuum in which the reflector
abundances scale together with self consistent line emission leads to
a strong increase in $\chi^2_\nu$ to $979/1152$ (i.e. $\Delta\chi^2$
of 6, compared to the fits in which the line normalisation is free)
for an incident continuum in which the abundances in the hot plasma
are all tied together except for Fe, and Si and S which are tied to
each other.  This is a strong indication that the iron abundance in
the {\it reflector} is too high with respect to the abundance of the
other elements.  The line is only consistent with that observed if the
iron abundance in the reflector is subsolar in its ratio with respect
to the other elements, and by an amount similar to that seen in the
hot plasma line emission i.e. $A_{\rm Fe}\sim 0.3$ while $A_{rest}\sim
0.5$.

\begin{table*}
\begin{minipage}{180mm}
\caption{Abundances with respect to Anders \& Grevasse (1989)}
\label{aoratag}
\begin{tabular}{ccccccccc}                \hline

kT\footnote{Maximum temperature in the plT models} (keV) & A (plasma)&
$\alpha$ & A (reflector) & $R$ & EW (eV) & 
$N_H$\footnote{$\times 10^{21}$ cm$^{-2}$} & $\xi$\footnote{Defined as
$L/nr^2$ where $r$ is distance from material of density $n$ illuminated
by a source of luminosity $L$} & $\chi^2_\nu$\\

\hline

$13.9^{+6.2}_{3.4}$ & Fe=$0.29^{+0.07}_{-0.05}$ & $0.15\pm 0.11$ & 
$0.5^{+0.75}_{-0.25}$ & $1.9\pm 0.9$ & $110\pm 30$ & $1.2^{+0.5}_{-0.3}$ & 
$0.18^{+0.30}_{-0.18}$ & 973.2/1150 \\
     & Si=$0.70^{+0.15}_{-0.13}$ & \\
     & S=$0.65^{+0.21}_{-0.18}$ & \\
     & rest=$0.40\pm 0.11$ & \\
\end{tabular}
\end{minipage}
\end{table*}

We replace the Anders \& Grevasse (1989) 'solar' abundances with those
of Morrison \& McCammon (1983), which has a factor 1.5 lower iron
abundance (see section 5.2), in both the plasma and reflection
codes. Details of some of these fits are given in Table 5.  We first
allow the line to scale independently of the reflection continuum, and
the reflection spectrum abundances to scale independently of the hot
plasma. With the abundance ratios tied at the (new) solar values we
get $\chi^2_\nu=1006/1153$ with $A\sim 0.4$. There is a marginally
significant change in the fit if the Fe abundance is fit separately,
with $\chi^2_\nu=1002/1152$. However, this is driven by the relative
overabundance of Si, and when Si is also free fit is significantly
better with $\chi^2_\nu=978.7/1151$ and the resulting Fe abundance is
consistent with that of the rest of the elements. There is a further
(marginally) significant decrease in $\chi^2_\nu$ to $973.6/1150$ with
S free to vary, but no other significant deviations from (new) solar
element ratios. In particular, allowing the abundance of O to be free
gives $\chi^2_\nu=973.4/1149$ for $A_O=0.42^{+0.20}_{-0.16}$.
The model with fewest parameters that describes the
data is then one in which Si and S are tied together, but are free to
vary from the Morrison \& McCammon (1983) solar ratios
($\chi^2_\nu=974.9/1152$, for $A_{Si, S}=0.65$, $A_{rest}=0.37$).  The
predicted iron line equivalent width with respect to the reflection
spectrum is then $\sim 0.7$ keV for $kT=14$ and $\alpha=0.15$. This is
a much better description of the fluorescent line seen in the data
than that expected from the
Anders \& Grevasse (1989) abundances. We set the reflector abundances
to be equal to $A_{rest}$, and include the self--consistent line
emission in the reflection spectrum, which gives a good fit at
$\chi^2_\nu=976.7/1154$.

\begin{table*}
\begin{minipage}{180mm}
\caption{Abundances with respect to Morrison \& McCammon (1983)}
\label{aoratmm}
\begin{tabular}{ccccccccc}                \hline
 
kT\footnote{Maximum temperature in the plT models} (keV) & A (plasma)
& $\alpha$ & A (reflector) & $R$ & EW (eV) &
$N_H$\footnote{$\times 10^{21}$ cm$^{-2}$} & $\xi$\footnote{Defined as
$L/nr^2$ where $r$ is distance from material of density $n$ illuminated
by a source of luminosity $L$} & $\chi^2_\nu$\\
\hline
$16.5^{+5.6}_{-4.3}$ & $0.42_{-0.06}^{+0.09}$ & $0.11\pm 0.11$ & 
$0.9^{+4.1}_{-0.5}$ & $1.5^{+0.9}_{-0.7}$ & $120\pm 40$ 
& $1.6\pm 0.6$ & $0.7^{+0.4}_{-0.5}$ & 1006/1153 \\
$13.9^{+6.2}_{-3.9}$ & Fe=$0.39^{+0.09}_{-0.08}$ & $0.17\pm 0.11$ & 
$0.5^{+1.1}_{-0.25}$ & $1.9\pm 0.9$ & $120\pm 40$ & $1.0_{-0.4}^{+0.6}$ & 
$0.36^{+0.90}_{-0.30}$ & 973.6/1150 \\
     & Si=$0.69^{+0.18}_{-0.16}$ & \\
     & S=$0.58^{+0.18}_{-0.17}$ & \\
     & rest=$0.38^{+0.11}_{-0.10}$ & \\

\end{tabular}
\end{minipage}
\end{table*}

\subsection{All Data}

We fit a power law temperature--emissivity plasma model with Morrison
\& McCammon (1983) abundances, together with its reflection spectrum
(also assuming these abundances) and a separate Gaussian line to all
the data simultaneously, with the abundances tied across all the
datasets. An ionized absorber is included for the ASCA data, and 
the plasma emission and Gaussian line normalisations, the maximum 
temperature and temperature weighting index are free to vary
between the spectra. This gives an excellent $\chi^2_\nu=998.5/1197$,
showing that the elemental abundances are consistent across all the
datasets, with $A_{Si, S}=0.64$ and $A_{rest}=0.40$. The two outburst
spectra are consistent with having the same
amount of reflection, with no increase in $\chi^2_\nu$ when the two
parameters are tied together, so that $\chi^2_\nu=998.5/1198$ for
$R=2.2$ in outburst and $R=0.7$ in quiescence. This difference
between the amount of reflection in outburst and quiescent is highly
significant: tieing all the reflected fractions together leads to an
increase in $\chi^2$ of 6.6. 

The best fit model spectra for the quiescent, GINGA decline from
outburst and ASCA outburst data fit simultaneously are shown in
Figures 4, 5 and 6, respectively. The plasma parameters are
$\alpha=3.9, 0.8, 0.19$, and $kT_{max}=13, 8.5, 12$, respectively, again
showing that most of the gas is hotter in quiescence than in outburst.

\begin{figure}
\plotone{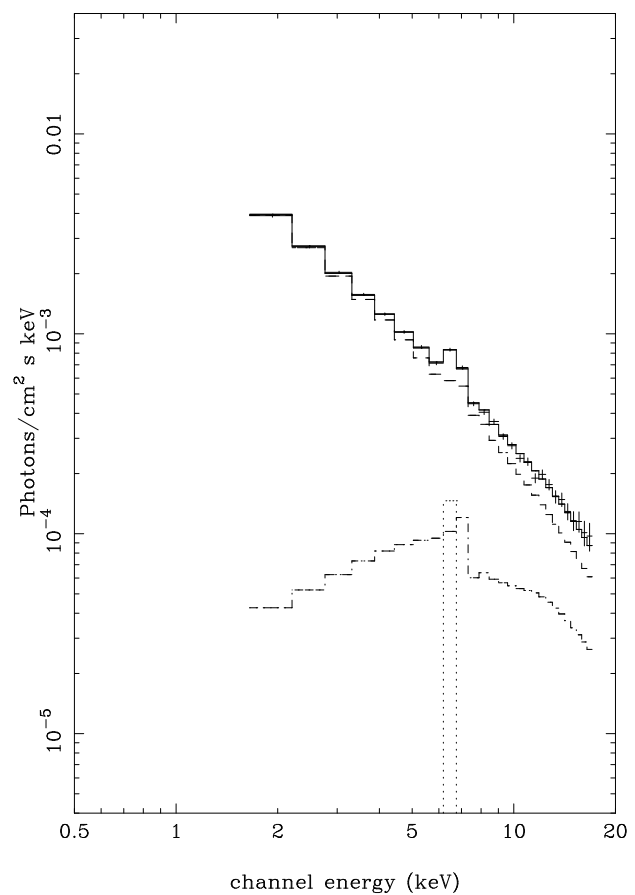}
\caption{The best fit GINGA quiescent spectrum.
}
\end{figure}

\begin{figure}
\plotone{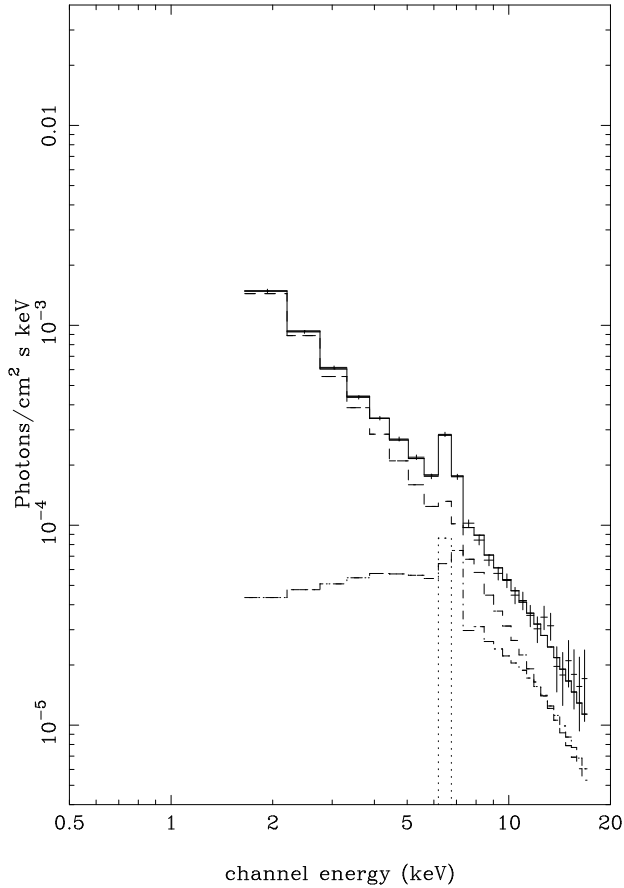}
\caption{The best fit GINGA decline from outburst spectrum.}
\end{figure}

\begin{figure}
\plotone{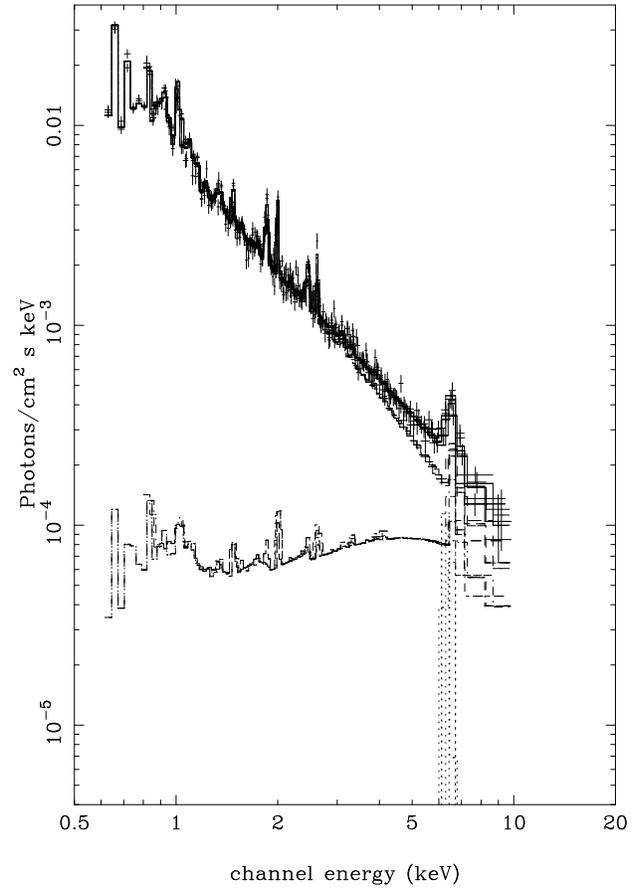}
\caption{The best fit ASCA anomalous outburst spectrum}
\end{figure}

\section{DISCUSSION}

All the data strongly require that the line emission is substantially
weaker than expected from a coronal plasma with solar abundances.
While this could simply indicate that the elements are subsolar, we
first examine alternative possibilities. 

\subsection{Validity of the Coronal Approximation}

The data have been fit with coronal plasma models, which assume that
radiative (and dielectronic) transitions are the dominant recombination
processes balancing collisional ionization.  All these processes
involve only one electron in a free--bound (or bound--free)
transition, so the resultant ion populations are independent of the
electron density to first order. 
Most of the ions are in the ground state as the
recombination rates are generally very much larger than the
collisional ionization rates.  However, at high densities collisional
excitation becomes important enough to rival radiative recombination,
so that the ions are generally in excited rather than ground
states.  Collisional ionization out of the {\it excited states}, a
process neglected in the assumption of a coronal plasma, 
can then modify the ionization balance. It also leads to 
suppression of the dielectronic recombination rate 
but this is not an important process at the temperatures considered here 
(e.g. Mewe 1990).  For Hydrogen--like ions with atomic number $Z$ in
a plasma of temperature $T$ (K), deviations from coronal ionization
balance are less than a factor 2 when the density $n\le 4\times 10^4 Z^2 T^2$
cm$^{-3}$ (e.g. Mewe 1990).  Typical temperatures where the
hydrogen--like state dominates for oxygen and iron are $10^7$ and
$10^8$ K, respectively, so these elements are out of coronal
equilibrium for densities $\ge 2\times 10^{20}$ and $\ge 2\times
10^{23}$ cm$^{-3}$.  The {\it lower} limit on the density inferred for
the boundary layer in the eclipse observations of HT Cas is $\ge
2.5\times 10^{13}$ cm$^{-3}$ (Mukai et al 1996). Thus there is no
direct evidence against coronal equilibrium approximations for the
ionization balance in the boundary layer in HT Cas in quiescence.
However, the mass accretion rate in SS Cyg in quiescence is nearly two
orders of magnitude larger than that in HT Cas, so its boundary layer
density may well be substantially higher. The models of Popham and
Narayan (1993) suggest a density of $\sim 10^{15}$ cm$^{-3}$ for
conditions appropriate to the quiescent GINGA spectrum. Again, in
outburst the boundary layer density is presumably higher still, but
the complexity of the flow makes it difficult to estimate the density
of the hard X--ray emitting material. Thus it seems unlikely that the
material is as dense as required for coronal ionization balance to
become invalid.

A generally more restrictive assumption in the coronal models is that
the material is optically thin. In HT Cas the eclipse measure can be
used to set constraints on the quiescent emission properties. Assuming
that the emission forms a corona of with outer radius 
$(1+\delta)R_{WD}$ then the
column density through the boundary layer is $N_H\sim 10^{13}
\delta^{1/2} R_{WD}\le 3\times 10^{21}$ cm$^{-2}$ for the maximum
value of $\delta=0.15$. This is optically thin to electron scattering
and is also likely to be thin to all the H and He--like Ly$\alpha$
transitions (Matt 1994; Mewe 1990). Even if these lines were optically
thick, there is no viable line destruction mechanism for these
transitions (see e.g. the discussion in Done et al 1995) so the
multiple scatterings have little effect.  The plasma is possibly
optically thick at iron L (Band et al 1990), but the fact that the
iron abundance as measured from the L lines agrees with that derived
from the K lines (see section 4.3) argues against this strongly
affecting the line emission. 

Self photo--ionization of the plasma is potentially important, since
there is an intense X--ray flux. The effective ionization parameter
for a diffuse source is $\xi=4\pi c U/n$ where $U\sim \epsilon \delta
R_{WD}/c$, and $\epsilon$ is the bremsstrahlung emissivity (see
e.g. Done et al 1995).  Thus $\xi\sim 1.8\times 10^{-14}n\sim 18$ for
a density of $\sim 10^{15}$ and $\delta=0.15$. This is not high enough
to disturb the ionization balance set by the collisional processes,
as the ionization states, even for Oxygen for a pure photoionized plasma
are much less than those produced in a collisionally ionized plasma
at $kT \sim 2-10$ keV (Kallman \& McCray 1982).

Another process that is neglected in the coronal models is three body
recombination (the inverse process of collisional ionization). Here a
free electron recombines with the ion, but the recombination energy is
given to another free electron rather than being radiated as a
free--bound transition. Since this involves two thermal electrons, the
rate depends on $n_e^2$ rather than $n_e$ as for the processes
discussed above, so it becomes important only at high densities. Again
it efficiently populates the upper excited levels of the ions,
so that collisional ionization from excited states can become
important and collisional excitation/de--excitation can change the
expected line strengths. Cota (1987), quoted by Rees, Netzer and Ferland
(1989), states that 3 body processes should be considered for all densities
$\ge 10^{10-11}$ cm$^{-3}$, which is clearly within the range of
densities which are important here. However, it is unclear
whether this can have an appreciable effect on the 2--1 line
transitions, and Keady et al (1990) show that the coronal
approximation for iron ion fractions in a 2 keV plasma holds up to densities 
$\sim 10^{20}$ cm$^{-3}$.

Coronal models also assume that the plasma is in equilibrium, and so
do not apply where the physical conditions vary on timescales which
are fast compared to the ionization/recombination timescales e.g. in
young supernovae remnants (see e.g. Shull 1982). The collisional ionization
timescale, $t_{ion}=(nC)^{-1}$, where $C$ is the collisional ionization 
rate coefficient, is slowest for H--like iron (Shull and Van
Steenberg 1982),  where $t_{ion}\sim 10^{16}/(nT^{0.5})$ at $\sim 10^8$ K,
i.e. 0.1 sec at $n=10^{13}$ cm$^{-3}$. 
This is much faster than the bremsstrahlung cooling timescale
$T_{cool}=1.5\times 10^{11} T^{0.5} n^{-1}$ sec for plasma at $10^8$ K, and
is even rapid compared to the observed X--ray variability (Yoshida et al 1992).
This it seems likely that the plasma is hot and dense enough to be
in ionization equilibrium.

One density dependent process that will clearly affect the 2--1 line
flux is collisional mixing of the 2s and 2p states. For an initially
bare nucleus, a recombination cascade can end with an electron
transition to the 1s ($1^2S$) level from either the 2s or 2p state
($2^2S$ or $2^2P$). The $2^2S$ level cannot decay down to ground via a
permitted electric dipole transition as it violates the selection rule
of $\Delta L=\pm 1$. Its only decay path is a 2 photon electric dipole
decay, leading to a continuum rather than an observable line. For
recombining initially H--like ions the situation is somewhat more
complex, as the cascading electron can have spin either aligned or
misaligned with that of the electron in the 1s state.  Thus an excited
2s--1s He--like ion can be either $^3S$, $^3P$, $^1S$ or $^1P$, of
which the $^1S$ state decays via 2 photons. Thus the K$\alpha$ line
from H and He--like ions can be increased by collisional mixing out of
the $2^2S$ and $2^1S$ level, respectively, and this can strengthen the
line by up to $50$\%, becoming effective for H--like ions of atomic
number $Z$ at densities greater than $n\sim 1.8\times 10^4 Z^{9.3}$
(e.g. Netzer 1996). Thus it should be important for O at densities
$\ge 5\times 10^{12}$ cm$^{-3}$, and all elements up to Si and perhaps
S for densities of $\sim 10^{15}$ cm$^{-3}$. However, it is unlikely
to affect iron as this requires densities $\ge 2\times 10^{17}$
cm$^{-3}$.  Thus in comparison to the plasma codes calculated with
$n=1$ cm$^{-3}$ we expect the observed Ly$\alpha$ lines from iron in
the hot plasma to be representative of the true abundance of iron,
while those from the lower $Z$ elements can be enhanced by this
process, leading to an overestimate of their abundance by a factor of
$\sim 1.2-1.5$.

In short, while true coronal equilibrium conditions are not fulfilled
in the boundary layer, none of the density, photo--ionization or
optical depth effects discussed above are likely to cause the observed
sub--solar abundance line intensities from the plasma. The one process
that should be important is {\it enhancement} of the Ly$\alpha$ lines
from H and He--like ions of low $Z$ elements by a factor $\sim
1.2-1.5$ by collisional mixing transforming states that decay via 2
photons into K$\alpha$ line. 

\subsection{Abundances}

Since coronal models seem to give a good indication of the X--ray line
equivalent widths, we can use these to estimate the abundances in SS
Cyg. If collisional mixing affects everything other than iron, then
all the elements are a factor $\sim 3$ sub--solar except Si and S,
which are probably more like a factor $\sim 2$ sub--solar. However, this
is {\it not} consistent with the intensity of the observed 6.4 keV
fluorescent line from the reflector in the ASCA data, which is weaker
in comparison to the reflection continuum than is expected from
material in which all the elements are subsolar by a factor 3. This
strongly suggests that the relative abundance of iron with respect to
the lighter elements is smaller than the 'solar' definition used
here. A recent abundance compilation by Grevasse et al., 
(1996) gives Fe/H of $\sim 3.2\times 10^{-5}$ compared to $\sim
4.6\times 10^{-5}$ in Anders \& Grevasse (1989), while most of the
lighter elements have abundances within 10\% of the Anders \& Grevasse
(1989) values.  Ironically, these newest abundances are very similar to 
those of Morrison \& McCammon (1983) for all the given elements. 
Without collisional mixing, the plasma in SS Cyg is then uniformly a factor
$\sim 0.4$ below the 'solar' values of Morrison \& McCammon (1983)
except for Si and S, which are $\sim 0.6\times$ 'solar'.

\subsection{Amount of Reflection}

The determination of the amount of reflection is dependent on the
assumed abundances. From the above discussion it seems that the
abundance ratios in the plasma and reflector are closer to those of
Morrison \& McCammon (1983). Using these abundances for the reflector
as well as the plasma gives the result for SS Cyg that $R=0.7$ in
quiescence, $R=2.2$ in outburst (see Section 4.4).  The amount of
reflection in quiescence is {\it not} consistent with the presence of
an optically thick inner disk (see Section 2.2 and 2.3). Either the
inner disk is disrupted, or is optically thin.  By contrast, the
larger amount of reflection in outburst requires that there is additional
reflecting material present compared to that in quiescence. 
This is consistent with a geometry in
which the boundary layer extends over the whole white dwarf surface,
illuminating both it and an inner disk.

\subsection{Ionized Wind}

A partially ionized absorber is seen in the ASCA outburst spectrum.
The derived ionization state for $\xi\sim 0.2-0.5$ (tables 4 and 5)
from the {\tt absori} code (see section 4.0) is such that the CIV
ionization fraction is at its maximum, with $f_{CIV}\sim 0.1$.  This
is the upper end of the range found for the blackbody boundary layer
or disk illumination models of Hoare and Drew (1993), where $f_{CIV}$
is typically $0.1-0.001$.  The X--ray column density in the warm
material in SS Cyg of $N_H\sim 10^{21}$ cm$^{-2}$ gives $\dot{M}\sim
2\times 10^{-12} M_\odot$ yr$^{-1}$ using the spherical wind velocity
law of Hoare and Drew (1993). Thus $\dot{M} f_{CIV}\sim 2\times
10^{-13} M_\odot$ yr$^{-1}$, much smaller than that inferred from the
UV resonance line modelling (e.g. Hoare and Drew 1993), perhaps
because the ASCA data are taken towards the end of an outburst (Nousek
et al 1993), where the wind may be declining rapidly. We stress that
the hard X--ray illumination does have important effects on the
ionization structure of the wind, and that the X--ray absorption
can give new constraints on the outflowing material which has hitherto
only been extensively studied in the UV.

\subsection{Continuum Models}

The lack of robust theoretical models for the hard X--ray spectrum is
a severe problem in interpreting the data. The calculations of Narayan
and Popham (1993) are the only numerical results for the quiescent
spectra that can be rigorously tested. However, these are restricted
by the 1 dimensional approximations which imply vertical averaging of
what is most likely to be a plasma with a stratified temperature
structure. In these models the expected spectrum is very like that
produced from material that is impulsively heated which then cools via
bremsstrahlung alone in a constant pressure/gravitational field. This
spectrum is {\it incompatible} with the GINGA quiescent spectrum,
in which the expected cooling gas is not present. Either these
models for the quiescent spectrum are wrong, and there is 
some additional process heating of the gas, or the cooler components
are masked by absorption. The ionized wind is 
is unlikely to be important in quiescence, so the only potential absorber is
an optically thin inner disk, which could attenuate the emission from 
the observable part of the hard X--ray boundary layer 
below the equatorial plane. However, this absorption is not
{\it required} by the data --  an equally acceptable description of the
spectrum is an X--ray source where the temperature--emissivity
relation is weighted towards higher temperatures than expected from
simple cooling models. 

In outburst there are no detailed models for the
temperature--emissivity structure that can currently be compared with
the data. However, the integrated luminosity from the hot gas is
similar in both outburst and quiescence, although its fractional
contribution to the bolometric luminosity is much lower in outburst.
The maximum temperature component of the emission is similar
in both quiescence and outburst at $\sim 10-15$
keV (see section 4.4). As expected, these are always less than the
maximum temperature of $\sim 20$ keV produced by a strong shock from
material accreting onto a white dwarf mass of $0.8M_\odot$ and radius
of $7\times 10^8$ cm (Frank et al 1992).

\section{CONCLUSIONS}

We find a significant reflection signature in all the SS Cyg
spectra. This is stronger relative to the illuminating plasma emission
in outburst than in quiescence, consistent with models in which the
inner disk is optically thick in outburst but optically thin or
non--existent in quiescence. The amount of reflection in both
quiescence and outburst also supports the idea that the hard X--ray plasma
forms a corona over the white dwarf surface, rather than being
confined to an equatorial band where the disk/star interaction takes
place.

The ASCA outburst spectrum shows significant absorption from partially
ionized material, which is probably the outflowing wind inferred from
the UV resonance line profiles. The X--ray absorption gives an
additional diagnostic window on the wind properties, and suggests
rather low mass loss rates of $\sim 10^{-12} M_\odot$ yr$^{-1}$.  The
hard X--ray emission strongly affects the ionization structure of the
wind, giving much larger CIV ion fractions than previous blackbody
boundary layer photoionisation models.

The ASCA data also give a good determination of the amount of line emission
from the plasma, which are a factor $\sim 2.5$ weaker than expected
for solar abundances. We critically examine the assumptions of coronal
equilibrium but conclude that the line strengths are representative of
the abundances.

The intrinsic plasma emission, when separated from the effects of
reflection and (in outburst) ionized absorption, can then be described
by hot plasma with a continuous (power law) temperature--emissivity
distribution.  In quiescence this distribution is strongly weighted
towards the highest temperatures, with very little cooler material
present, so that the emission can in practice be modeled by a single
temperature plasma. The lack of cool components is in conflict with
the predictions of the advective boundary layer models of Narayan \&
Popham (1993), and with simple models of cooling gas! One possible way
to reinstate these models is for the quiescent spectrum to be
distorted by partial absorption, such as would be produced if the
inner disk is optically thin to electron scattering but optically
thick to photoelectric absorption. The flux from the boundary layer
below the disk is then absorbed, while that from above the disk is
not.

The outburst spectrum by contrast is much softer. It is dominated by
the cool components and cannot be modeled as single temperature
emission, but the total luminosity of the X--ray emitting gas is very 
similar to that in quiescence. 
We review current theoretical models for the hard X--rays
in outburst, but none of these address the probable difference in the
physical nature of the viscosity between the disk and the boundary
layer, or the inherently 3 dimensional nature of the problem. 
We urge further theoretical modelling, giving the predicted,
spatially integrated temperature--emissivity distribution, so that
a better understanding of the spectra of dwarf novae can be developed. 

\section{ACKNOWLEDGEMENTS}

We thank Dave Smith for his help with the GINGA data extraction, Bob
Popham for discussions about the advective boundary layer models, and
John Parker for spatial perception in the reflection calculations. CD
acknowledges support from a PPARC Advanced Fellowship. This research
has made use of data obtained through the High Energy Astrophysics
Science Archive research Center Online Service, provided by the
NASA/Goddard Space Flight Center, and from the Leicester Database and
Archive Service at the Department of Physics and Astronomy, Leicester
University, UK.

\section{APPENDIX}

The angle $\eta$ between the X--ray emitting source and the line of sight 
is then given by 
$$\cos \eta = \sin i \sin \theta \sin \phi + \cos i \cos \theta$$.

The reflection spectrum from each small segment of the white dwarf underlying the boundary 
layer can be approximated by an isotropically illuminated slab
viewed at inclination $\eta$. Thus the average inclination between the 
reflector and the observer is 
given by 
averaging $\eta$ over all angles at which the boundary layer emission can be seen.
If the disk is opticaly thick and extends down to the white dwarf surface (figure 2a)
then this is given by

\begin{eqnarray}
\overline{ \eta(i,\beta)} & = & {2\over N}
\int_0^{\pi/2} d\phi \int_{\pi/2-\beta}^{\pi/2}
d\theta \eta(\theta,\phi,i) \sin\theta \nonumber \\
&   & +{2\over N}\int_{-\phi_{\rm min}}^{0} d\phi \int_{\pi/2-\beta}^{\theta_{\rm max}}
d\theta \eta(\theta,\phi,i) \sin\theta  
\end{eqnarray}

\noindent where the normalisation $N$ is given by equation (1)
but with all the $\eta(\theta,\phi,i)$
terms removed from the integrand, 

\noindent and 
\begin{eqnarray}
\phi_{\rm min} & = & \pi/2 \ldots \beta\ge i \nonumber \\
& = & \arcsin \Bigl[{\tan\beta \over \tan i } \Bigr] \ldots \beta\le i \nonumber \\
\theta_{max} & = & \arctan \Bigl( {-1\over \sin\phi \tan i} \Bigr) \nonumber
\end{eqnarray}

This needs to be integrated numerically. The inclination angle
of SS Cyg is thought to be  $i=37^\circ$ (Cowley et al 1980).
For the two limiting values of $\beta\to 0^\circ$ and $\beta\to 90^\circ$ 
the angle of the emission to the line of sight is $67^\circ$ and 
$53^\circ$, respectively. 

Alternatively, if the inner disk does not obscure the boundary layer (figure 2b)
then the mean inclination angle is 
\begin{eqnarray}
\overline{ \eta(i,\beta)} & = & {2\over N}
\int_{\phi_{\rm min}}^{\pi/2} d\phi \int_{\pi/2-\beta}^{\pi/2+\beta}
d\theta \eta(\theta,\phi,i) \sin\theta \nonumber \\
& & + {2\over N} \int_0^{\phi_{\rm min}} d\phi \int_{\pi/2-\beta}^
{\pi+ {\theta_{\rm max}} } d\theta \eta(\theta,\phi,i) \sin\theta \nonumber \\
& & + {2\over N}\int_{-\phi_{\rm min}}^0 d\phi \int_{\pi/2-\beta}^{\theta_{\rm max}}
d\theta \eta(\theta,\phi,i) \sin\theta
\end{eqnarray}
 
\noindent where $\theta_{\rm max}$ and $\phi_{\rm min}$ are given as before, and
the normalisation $N$ is given by equation (2) with all the $\eta(\theta,\phi,i)$
terms removed from the integrand.

Again, the limiting case for $\beta\to 0$ is a mean inclination of $53^\circ$
(the presence or absence of a disk makes no difference if the boundary layer
is limited to the equatorial plane), while that for $\beta\to 90^\circ$
is $57^\circ$.

Given that all these limits are very close to each other, 
the mean angle at which we see the boundary layer in SS Cyg is
approximately $60^\circ$, irrespective of whether or not there is an 
inner disk or the extent of the boundary layer.
Thus the reflection component from the white dwarf surface in SS Cyg is
approximately that expected from an isotropically illuminated slab viewed at 
$60^\circ$.

\end{document}